\documentstyle[epsf,11pt]{article}
\newcommand{\bea}{\begin{eqnarray}}
\newcommand{\eea}{\end{eqnarray}}
\begin{document}
\vspace{0.2cm}

\newcommand{\lsim}
{{\;\raise0.3ex\hbox{$<$\kern-0.75em\raise-1.1ex\hbox{$\sim$}}\;}}
\newcommand{\gsim}
{{\;\raise0.3ex\hbox{$>$\kern-0.75em\raise-1.1ex\hbox{$\sim$}}\;}}

\begin{flushright}
{\large HIP-2003-60/TH \\[2mm]} 
\end{flushright}

\begin{center}
{\LARGE\bf Single sneutrino production at hadron colliders}
\\[15mm]
{\bf Masud Chaichian$^{a,b}$, Anindya Datta$^{b,}$\footnote{Present Address
: INFN, Sezione di Roma, Universita La-Sapienza, P.le A. Moro 2, I-00185,
Rome, Italy}, Katri Huitu$^{a,b}$, 
Sourov Roy$^{b,c,}$\footnote{Present address: Helsinki Institute of Physics}, and Zenghui Yu$^{a,b}$}\\[4mm]
$^a$Division of High Energy Physics, Department of Physical Sciences, and \\
$^b$Helsinki Institute of Physics, \\
P.O.Box 64, FIN-00014 University of  Helsinki, Finland \\[4mm]
$^c$Department of Physics, Technion - Israel Institute of Technology \\ 
Haifa 32000, Israel \\[7mm]
\date{}
\end{center}

\begin{abstract}
We study the production of a single sneutrino in association with one
or two $b$-quarks at hadron colliders, in the framework of
an R-parity violating supersymmetric model. We find that at the Large
Hadron Collider (LHC) four $b$ final
states are promising with efficient $b$-tagging.  $l^+l'^-$ decay modes
of the sneutrino can also be viable for detection at the
LHC. However, the branching ratio for rare $\gamma\gamma$ decay
channel is too small to be seen.
\end{abstract}

\vskip 10mm

\baselineskip=0.2in

\noindent
Detection of supersymmetric (SUSY) particles will be one of the major
goals at the LHC. The common strategy for the detection is to study
the production and cascade decays of strongly interacting
superparticles like squarks and gluinos.  The production
cross-sections for sleptons or sneutrinos are small as they are weakly
interacting.  At the same time, pair production of these weakly
interacting superpartners may not be favoured by kinematics even at
the LHC.  Single production of these particles might be useful to
consider in such situations.

The supersymmetric partners can be produced singly, if the R-parity
\cite{RP_Farrar} is broken ($R_{p}=(-1)^{3B+L+2S}$, where $B$, $L$ and
$S$ denote the baryon number, lepton number and spin, respectively).
Another important consequence of the $R_{p}$-violation ($\rlap/\!
R_{p}$) is that it may explain the experimental results on neutrino
masses from atmospheric \cite{neutrino1}, solar neutrinos
\cite{SNO,neutrino2} and reactor \cite{kamland} experiments.  In the
sneutrino sector, sneutrinos and antisneutrinos can mix, which can
give rise to CP violation \cite{bes2}.  Detection of weakly
interacting, neutral sneutrinos at hadron colliders may not be
straightforward, but since they may be among the lightest
supersymmetric particles with interesting properties, it is important
to explore all possibilities for their detection.

We will consider here the $L$-violating trilinear terms \cite{RP_V_form}:
\bea
    W_{\rlap/\! R_{p}} & =\frac{1}{2}
\lambda_{[ij]k} L_{i}.L_{j}\bar{E}_{k}+\lambda^{'}_{ijk}
L_{i}.Q_{j}{\bar D_{k}},
\eea
where $L_i$ and $Q_i$ are the SU(2) doublets containing lepton and quark
superfields, respectively, $\bar{E}_j$ ($\bar{D}_j$, $\bar{U}_j$)
are the singlets of lepton (down-quark and up-quark),
and $i,j,k$ are generation indices and square brackets on them denote
antisymmetry in the bracketed indices.

The single sneutrino production at hadron colliders has
been studied in Refs. \cite{HR,krsz,drs,mpp}. 
The resonant sneutrino produced in the Drell-Yan process via first
generation $\lambda'$ type couplings could also decay via the same
couplings, but its detection seems problematic due to the large QCD
background \cite{HR} especially at LHC.
Therefore, other decay channels following the resonant production have been
studied, {\it e.g.} decays to a lepton pair \cite{krsz} and decays via
gauge interactions \cite{drs}, including cascade decay
to three leptons \cite{mpp}.
In all these cases the detection seems viable for some part of the 
parameter space.
In Ref. \cite{bews} the rare decay mode to two photons was considered.
It was found that taking into account single sneutrino production from
several different two parton processes, the accumulated events may
provide a sufficient signal at the LHC, at least for relatively light 
sneutrino mass and large $\rlap/\! R_{p}$ coupling.
The single production of sneutrinos in lepton colliders has been 
considered in \cite{RP_spp}.

We will assume here that the $\rlap/\! R_{p}$ couplings have a 
family hierarchy and only the third generation  fermions couple
significantly to the sneutrinos.
Thus the valence quark contribution to the parton level Drell-Yan 
process is small.
\begin{figure}[t]
%\vspace*{-2.5truecm}
\centerline{
\mbox{\epsfxsize=11.truecm\epsfysize=5.truecm\epsfbox{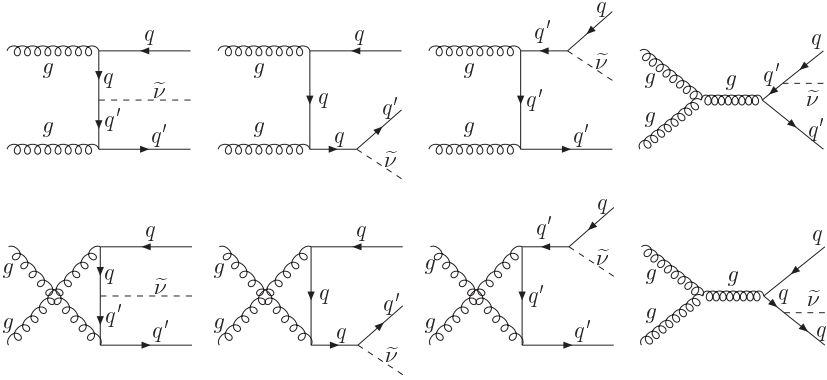}}
}
%\vspace*{-10cm}
\caption{\label{fig1}Some of the parton level
Feynman diagrams contributing to the process 
$pp \rightarrow \tilde{\nu}qq'$.}
\end{figure}
Here we will concentrate on the single production of sneutrinos 
in proton proton (anti-proton) process
in association with two jets, of which at least one is a $b$-quark.
Efficient $b$-tagging helps to find out the signal events.
Some of the Feynman graphs are depicted in Fig. \ref{fig1}.
There are two classes of sub-processes (quark initiated and gluon
initiated) contributing to the total cross-section.  At the LHC, gluon
fluxes are larger than quark fluxes. In addition, in the quark
initiated sub-processes one of the initial partons is an anti-quark,
and thus is excited from the sea. Eventually from a proton-proton collision,
effective $q \bar q$ luminosity is negligible compared to the gluon-gluon
luminosity at the LHC. At the Tevatron, which is a proton anti-proton
collider, both the initial quark and anti-quark can be valence
partons. 
Although the gluon initiated contribution is dominant also here, the quark
initiated subprocesses are not negligible.
We can easily see this by calculating the ratio of the cross section 
for quark initiated contributions to the
cross section, in which both quark and gluon initiated contributions
are included.
At the LHC for a sneutrino of mass 180 GeV (with $\lambda ' = 1$), 
the ratio is nearly 1/135 (1.35 pb) compared to 1/5 (3.5 fb)
at the Tevatron.

The production of single sneutrinos 
depends only on two unknown parameters, namely the sneutrino mass and the 
value of the relevant coupling $\lambda'_{ijk}$.

The constraints on $R_p$-violation coming from the low energy
experiments have been widely discussed \cite{RP_Phe}.
Here we will consider those third generation couplings which are
relevant for us.
We will first discuss the bounds other than those due to the
neutrino mass.
The best bound on $\lambda'_{33k}$ comes from the ratio
$R_l=\Gamma (Z\rightarrow had)/\Gamma (Z\rightarrow l\bar l)$.
Assuming that the $\tilde q$ masses in the loop are above
200 GeV, the upper bound becomes $\lambda'_{33k}\lsim 0.55$ \cite{bes}.
For us the other relevant experimental limits on the couplings are 
(from Allanach et al. in \cite{RP_Phe})
\bea
&&\lambda_{32k}\lsim 0.070\times \frac{m_{\tilde e_{kR}}}{100 \;{\rm GeV}},
\;
\lambda^{'}_{323}\lsim 0.52\times \frac{m_{\tilde b_R}}{100 \;{\rm
GeV}} . 
\label{bounds}
\eea
The bounds in (\ref{bounds}) are found \cite{bgh} from the measurements of
$R_\tau=\Gamma (\tau\rightarrow e\nu\bar\nu)/
\Gamma (\tau\rightarrow \mu\nu\bar\nu)$ and 
$R_{\tau\mu}=\Gamma (\tau\rightarrow \mu\nu\bar\nu)/
\Gamma (\mu\rightarrow e\nu\bar\nu)$ for $\lambda_{32k}$, and
from \cite{ls} 
$R_{D_s}=\Gamma (D_s\rightarrow \tau\nu_\tau)/
\Gamma (D_s\rightarrow \mu\nu_\mu)$ for $\lambda'_{323}$.

It has been shown that phenomenologically acceptable neutrino
masses can be generated by using only $R$-parity violation, see
e.g. \cite{ABL}.
The fit results in $R$-parity violating couplings, which are quite
small in size.
We will not insist in generation of the neutrino mass matrix, but
instead take a more conservative viewpoint, and take into account
only the limits in the previous paragraph, and consider the 
possibilities to detect a sneutrino at LHC.

The mass bounds for the sneutrinos and other supersymmetric particles
in $R$-parity violating models have been studied in the LEP and
Tevatron experiments \cite{searches,aleph}.
In \cite{aleph} it was found that if the $\lambda'_{3jk}$ coupling
dominates,
the lower limit on the tau sneutrino mass varies between
79 GeV and 92 GeV, depending on the lightest neutralino mass.
If the lightest neutralino is heavier than around 80 GeV, the lower
bound on the sneutrino mass drops to slightly above 40 GeV.

\begin{figure}[t]
\vspace*{-3truecm}
\centerline{
\mbox{\epsfxsize=7.5truecm\epsfysize=10.5truecm\epsffile{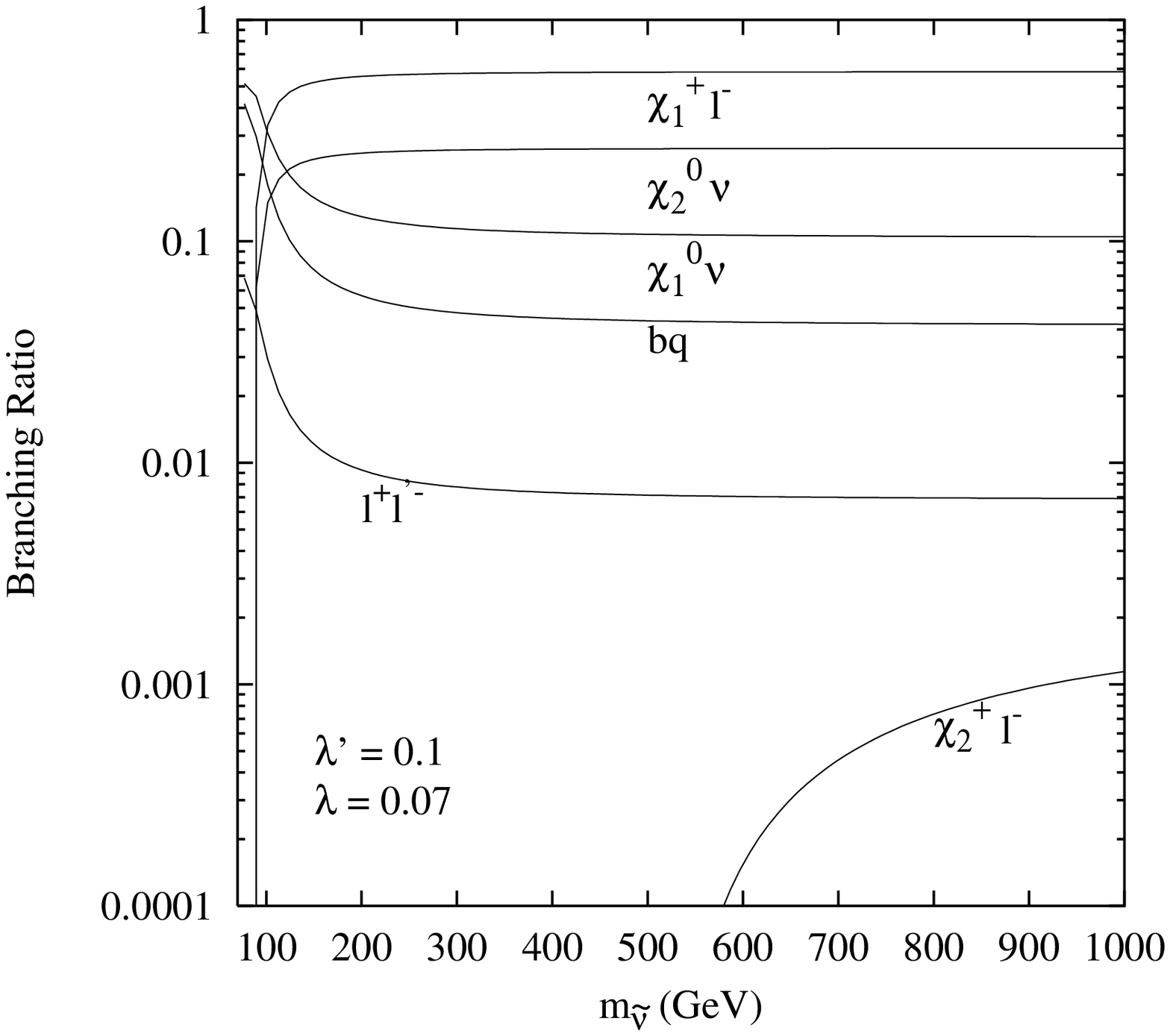}}
\mbox{\epsfxsize=7.5truecm\epsfysize=10.5truecm\epsffile{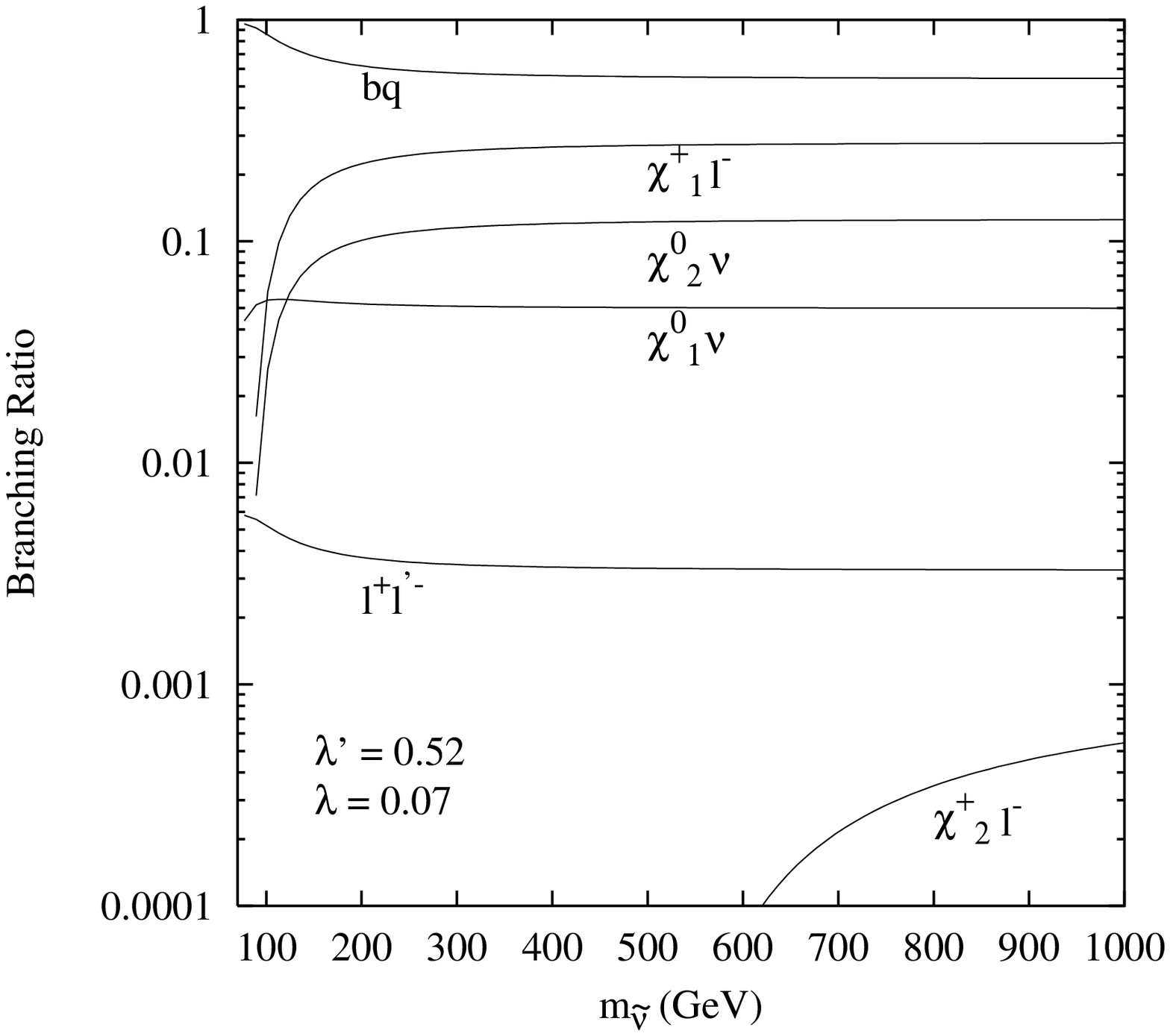}}
}
\caption{\label{BR}
 The branching ratio of a sneutrino, when $\lambda'=0.1$ or 0.52,
 and $\lambda=0.07$. In both the cases $\mu$ = 500 GeV and $\tan\beta$ = 10.}
\end{figure}
The $\tilde\nu$ decay modes are essential for the detection.
These depend on the SUSY parameters.
Here we will consider the sneutrino decay modes to at least one 
$b$-jet, leptons
and photons, {\it i.e.}
\bea
\tilde\nu  \rightarrow  bq,\;  l^+l'^-,\;   \gamma\gamma .
\eea
In our example of the branching ratios for the sneutrino in 
Fig. \ref{BR}, the lightest
neutralino mass is 41 GeV, the second lightest neutralino and the
lightest chargino $\sim 80$ GeV and the heavy chargino $\sim 500$ GeV.
The value of $\mu$ is taken to be 500 GeV and $\tan\beta$ = 10.
With the coupling $\lambda'=0.52$, the decay channel to two quarks
is the dominant one.
A heavier neutralino or chargino mass would significantly increase the 
R-parity violating branching ratio of the sneutrino. 
This in turn implies a higher rate
of the sneutrino decay signal we are interested in. 
The decays via $\lambda'$ interactions do not require any new
information of the model, since the decay can occur via the same
coupling as the production.
The other relevant decay modes are the gauge decays, which are
dominant, if the  $\rlap/\! R_{p}$ couplings are small.
The decays via gauge interactions lead typically to complicated
cascade decays \cite{drs}.
The branching ratio to two photons is ${\cal{O}}(10^{-6})$.
In Fig. \ref{BR}, we have also assumed that $\lambda=0.07$, and
consequently the sneutrino decays to two leptons as well.

\begin{figure}[t]
\vspace*{-3truecm}
\centerline{
\mbox{\epsfxsize=8.truecm\epsfysize=11.truecm\epsffile{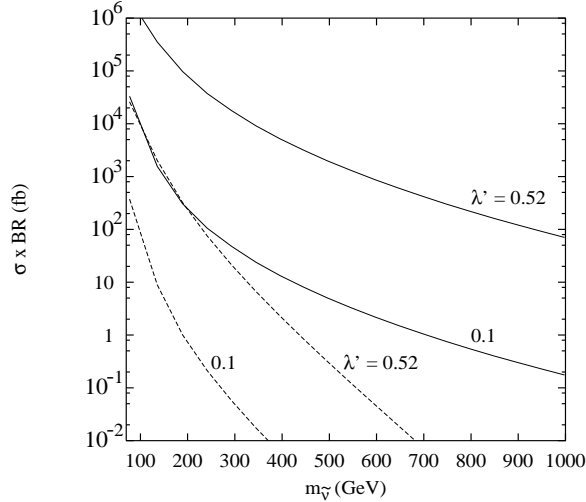}}
}
\caption{\label{bb}
The cross section for the process
$pp \rightarrow bq\tilde\nu\rightarrow
bq bq$ as a 
function of the sneutrino mass with
$\lambda^{'}$
=0.52, 0.1, as
indicated in the figure.
The solid lines correspond to LHC and the dashed ones to Tevatron.}
\end{figure}
In our numerical calculations we use for the gluon distribution 
functions the PDFLIB package \cite{pdflib} and the GRV distribution 
\cite{GRV} from there.

When one of the couplings $\lambda'_{323}$, $\lambda'_{332}$, or 
$\lambda'_{333}$ does not vanish, 
we shall have two or four $b$-jets in the final state.
If only one of the couplings $\lambda'_{323}$ and $\lambda'_{332}$ 
is different from zero, 
in the final state one observes two $b$-jets and two $s$-jets 
with a pair of a $b$-jet and an $s$-jet
forming the sneutrino, while if also
$\lambda'_{333}\ne 0$, one can have final states with three or
four $b$-jets.
In Fig. \ref{bb} we have plotted the production cross section times
the branching ratio for LHC (solid) and Tevatron (dashed) using two
values for the coupling, $\lambda'=0.52$ and 0.1.
For LHC the cross sections remain above 1 fb up to $m_{\tilde\nu}\sim
700$ GeV, while for Tevatron the cross sections are considerably 
lower.

%-----------------------------------------------------------------
\begin{figure}[t]
\leavevmode
\begin{center} 
\mbox{\epsfxsize=8.truecm\epsfysize=8.truecm\epsffile{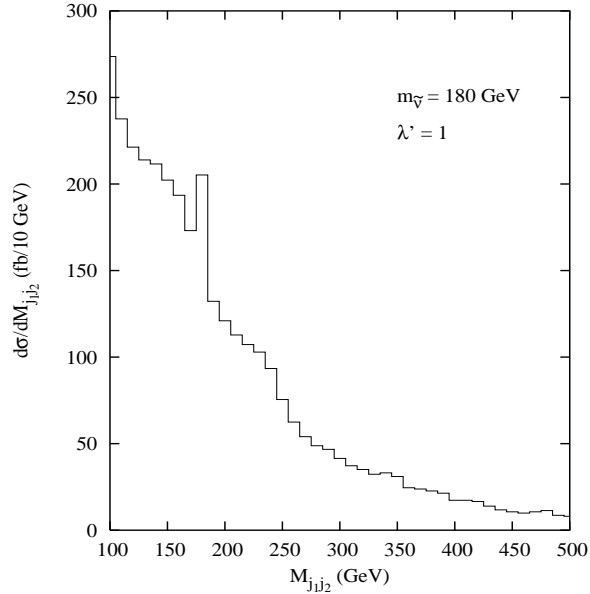}}
\end{center}
\caption{\label{pt_dist}
Invariant mass distribution of two highest $p_T$ jets, of sneutrino events 
superimposed on SM QCD background. We have used bin sizes of 10 GeV. 
For the sneutrino signal a sneutrino mass of 180 GeV and $\lambda ' =$  1 
have been used.} 
\end{figure}
%-------------------------------------------------------------------

The signal that we are interested in, is not free from the Standard
Model (SM) backgrounds.  In fact, the cross-section of the SM
processes contributing to the four jet final state is huge.  A large
number of diagrams contribute to this final state. We have estimated
this SM background using the package MADGRAPH \cite{madgraph} and
HELAS \cite{helas}. The SM QCD background is populated mostly at the
low transverse momentum and high rapidity of the jets.  We have
demanded that the final state is comprised of exactly four jets with
rapidity $\vert \eta \vert < 2.5$ and transverse momentum $p_T > 25$
GeV. We also demand that the angular separation in between any two of
the jets is substantial, $\Delta R \;\;(= \sqrt{(\Delta \eta)^2 +(\Delta
\phi)^2 }) > 0.7$.   Using only these cuts does not help us to reduce
the SM background much.  We notice that the jets coming from sneutrino
decay have larger transverse energy/momentum. This is demonstrated in
Fig. \ref{pt_dist}, where we have plotted the invariant mass
distribution of two highest $p_T$-jets for signal and background for a
180 GeV sneutrino mass. For the signal, invariant mass distribution of
two highest $p_T$-jets peaks sharply (modulo the detector smearing and
decay width of the sneutrino) around the sneutrino mass compared to
the monotonically decreasing distribution from SM QCD processes. We
have taken into account the finite detector resolution effects by
gaussian smearing of the $p_T$ of the jets, as $\Delta p_T^j / p_T^j =
0.6/\sqrt{p_T^j} + 0.03$. 
Thus we compare the number of signal and background events in the bin
(of width 10 GeV) corresponding to the
sneutrino mass.

There is one caveat regarding the normalisation of the SM
background. The background cross-section is proportional to the fourth
power of $\alpha_s$. As we are only using the leading order (in
$\alpha_s$) expression for the background cross-section, the scale
dependence of the result is quite strong. To be conservative, we have
chosen the scale of $\alpha_s$ (and the factorisation scale of parton
distribution function) to be equal to the $p_T$ of the softest
jet. While estimating the sneutrino cross-section, the same scale is
set to be equal to $\sqrt{\hat s}$, center-of-mass energy of the
colliding partons. In this way, we have tried to maximise the
background and minimise the signal, to make a conservative estimate 
of signal to background ratio.

%-----------------------------------------------------------------
\begin{figure}[t]
\leavevmode
\begin{center} 
\mbox{\epsfxsize=7.truecm\epsfysize=7.truecm\epsffile{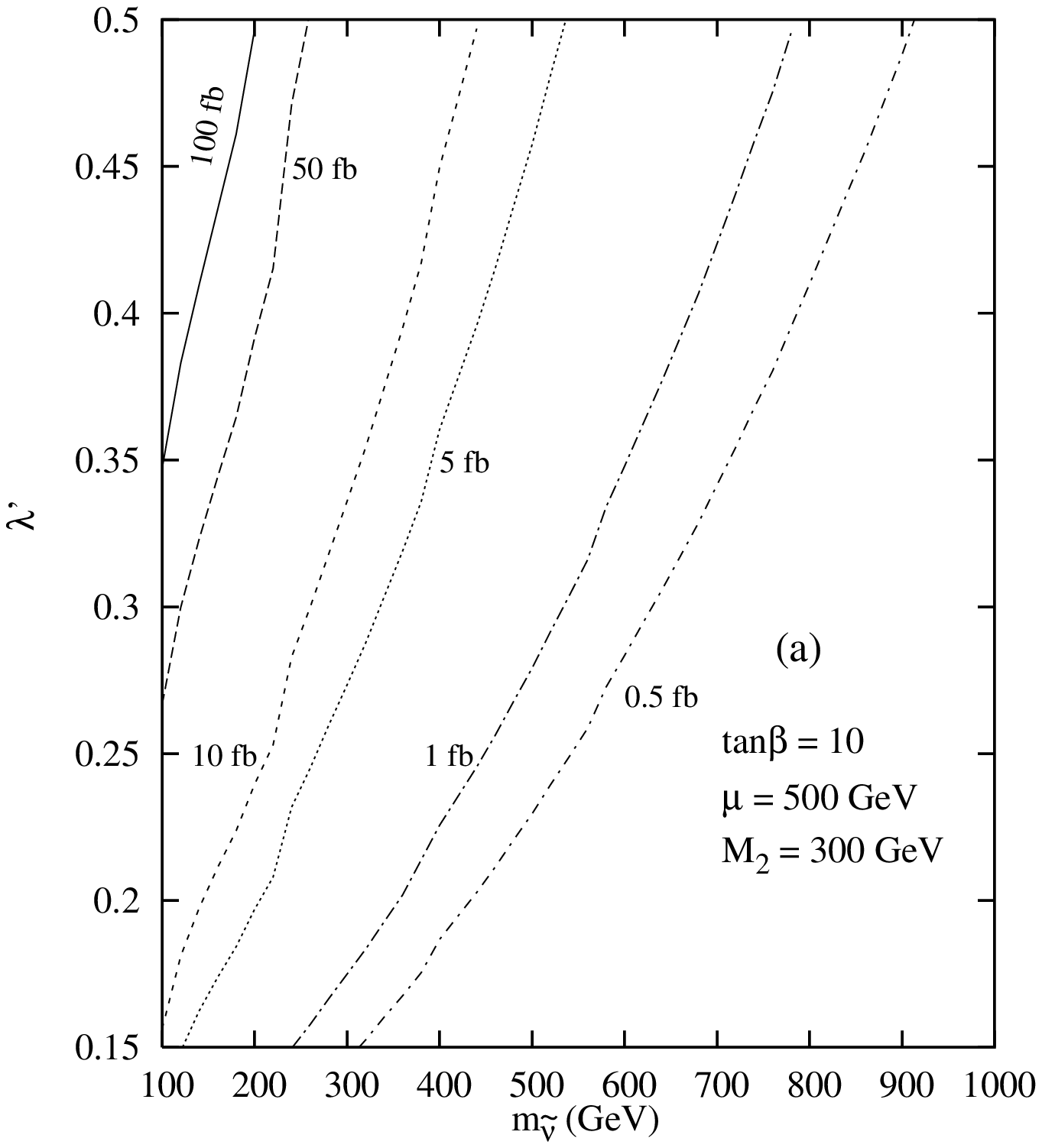}
\epsfxsize=7.truecm\epsfysize=7.truecm\epsffile{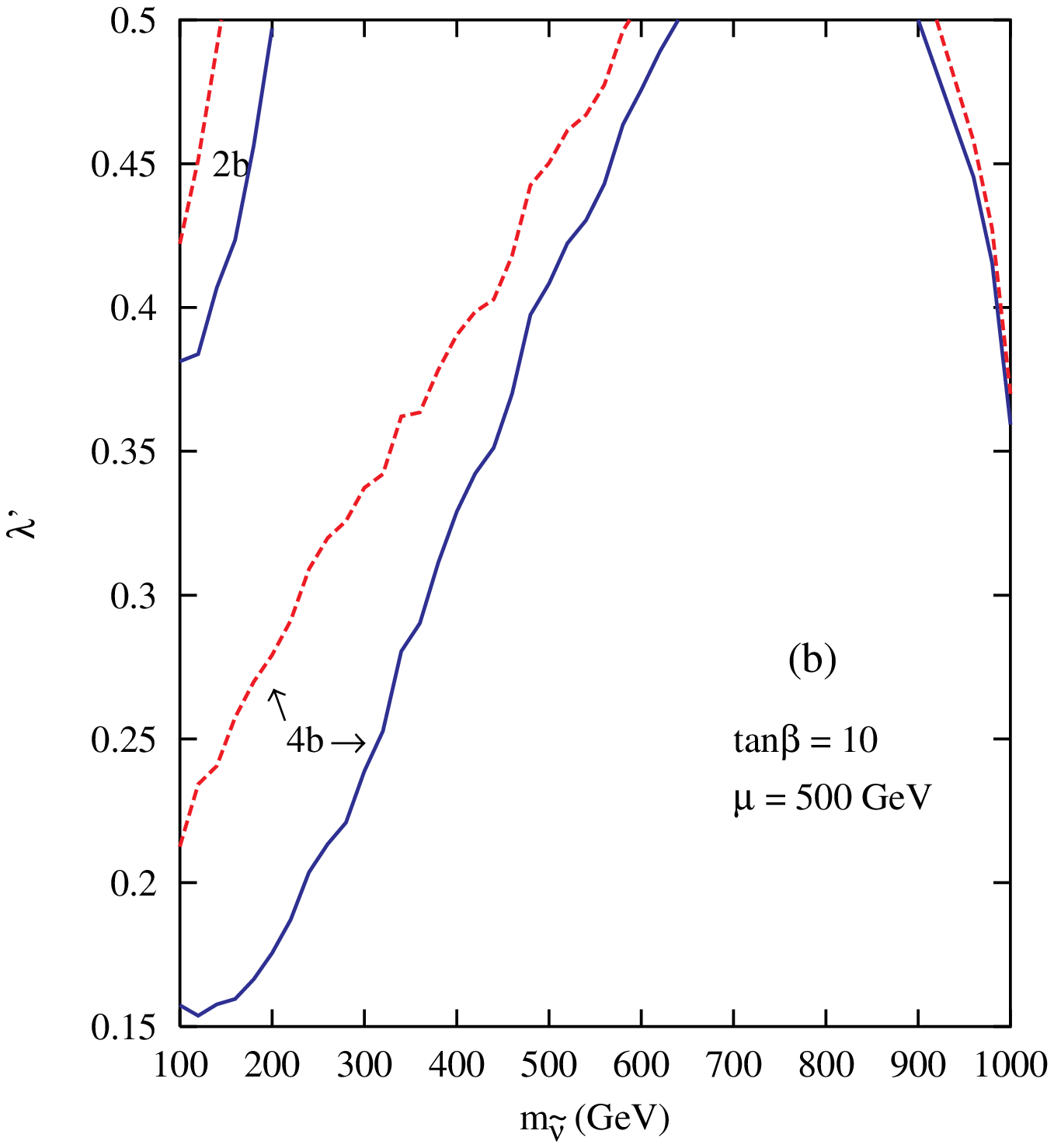}
}
\end{center}
\caption{\label{reach}
a) Contour levels of $4b$-cross-section in invariant mass bins (of
width 10 GeV, central value of the bin corresponds to sneutrino mass)
from sneutrino production and decay via $\lambda '$ coupling. We have
used $\tan \beta = 10$, $\mu = 500$~GeV and $M_2 = 285$~GeV for this
plot.
b) 5$\sigma$ discovery regions for LHC (with 100 fb $^-1$
integrated luminosity) in $\lambda ' - m_{\tilde \nu}$ plane for
$m_{\tilde\chi_1^\pm}=81$ GeV (dashed) and $m_{\tilde\chi_1^\pm}=272$ 
GeV (solid).  Lines marked with `2b' represent the final states containing
two $b-$jets and lines marked with `4b' represent final states with
four $b$-jets.}
\end{figure}
%-------------------------------------------------------------------

In Fig. \ref{reach}a, the contour levels for four $b$-quark cross-section
in bins (of invariant mass of two highest $p_T$ jets) of 10 GeV have been
presented for $\tan \beta = 10$, $\mu = 500$~GeV and $M_2 = 285$~GeV. 
The central value of the bin corresponds to the sneutrino
mass.  The nature of the contours can be easily explained. Sneutrino
cross-section falls off with mass, and to make up this decrement one has
to have higher values of $\lambda'$ coupling. The mild 'knees' around
$m_{\tilde \nu}$ = 250 GeV and 550 GeV are due to the opening up of the 
sneutrino decay channels to the second lightest neutralino, lighter chargino, 
and heavier chargino.

In Fig. \ref{reach}b, we have plotted the statistical significance
($\sigma \equiv$ signal/$\sqrt{\rm background}$) of the proposed signal over
SM background with 100 fb$^{-1}$ luminosity at the LHC. We have
considered two different cases of R-parity violating couplings. The
lines marked with '4b', assume non-zero values for $\lambda ' _{i33}$,
and represent thus final states with four $b$-jets. The curves marked
with `2b' represent the final states of two $b$- and two light quark
jets (gluons in the case of SM background). 

Since no sfermion mixing is involved here our results are not very
sensitive to $\tan\beta$. However, relative strength of $\mu$ with
respect to $M_2$, can alter the compositions of charginos and
neutralinos. For presentation we have used $\mu$ = 500 GeV with $M_2$
= 85 GeV (correspond to the chargino mass $m_{\tilde\chi_1^\pm}=81$
GeV; dashed line). This configuration essentially results in a
wino-like lighter chargino and second lightest neutralino along with a
bino-type lightest neutralino. Values of $\mu$ comparable with $M_2$
(285 GeV, solid line) could result in the lightest neutralino and
charginos with competing gaugino and higgsino components. This value
of $M_2$ corresponds to $m_{\tilde\chi_1^\pm}=272$ GeV. Larger higgsino
components in the lightest neutralino and lighter chargino will
diminish the sneutrino decay rates to R-parity conserving
channel. This is evident from the nature of the plots presented above.
The value of $\mu$ (relative to $M_2$ values) used in our
analysis is on the conservative side.

In the regions above the solid and dashed lines, signal strength is
higher than a $5\sigma$ fluctuation of the SM background.  It is
evident from the figures that discovery reach for the '4b' channel is
better than for the '2b' channel where sneutrino is decaying to a $b$-
and a non-$b$-quark. This can be accounted for by the size of SM
background in these two cases. With an efficient $b$-tagging (as
assumed in our case), '4b' background is more under control compared
to the '2b' + '2j' final state.
It has
been assumed in the calculation that the $b$-tagging efficiency is 60
$\%$, and mis-tagging probability is 1 $\%$ \cite{btag}.  For Tevatron
the $5\sigma$ effect can be found in the case of the four $b$-jets,
when the sneutrino mass is around 100 GeV and $\lambda'\sim 0.5$.

Both the signal and background cross-sections decrease monotonically
with increasing invariant mass of the two highest $p_T$ jets (equals
to sneutrino mass for the signal). However, sneutrino cross-section
falls off more rapidly than the QCD background for lower values of the
invariant mass. Thus to have a constant value of statistical
significance for heavier sneutrino masses we need higher value for
$\lambda'$. This is evident from the plots in
Fig. \ref{reach}. However, beyond some particular value of this
invariant mass, the situation is reversed and QCD cross-section
decreases more rapidly than the sneutrino cross-section. This explains
the behaviour of the plots ($4b$-case only) for higher values of
invariant mass. We have estimated the prospect of our proposed signal
for two different values of the chargino (neutralino) mass. For a
higher value of chargino mass (272 GeV in Fig. \ref{reach}),
the R-conserving sneutrino decays are suppressed resulting in an
enhancement of our signal coming from R-violating channel. Naturally
efficacy of the R-parity violating decay signal for a higher value of
chargino (neutralino) mass is better than a lower chargino mass (for
illustration we have also presented the result for a chargino mass of 81
GeV).

\begin{figure}[t]
\vspace*{-3truecm}
\centerline{
\mbox{\epsfxsize=7.truecm\epsfysize=9.5truecm\epsffile{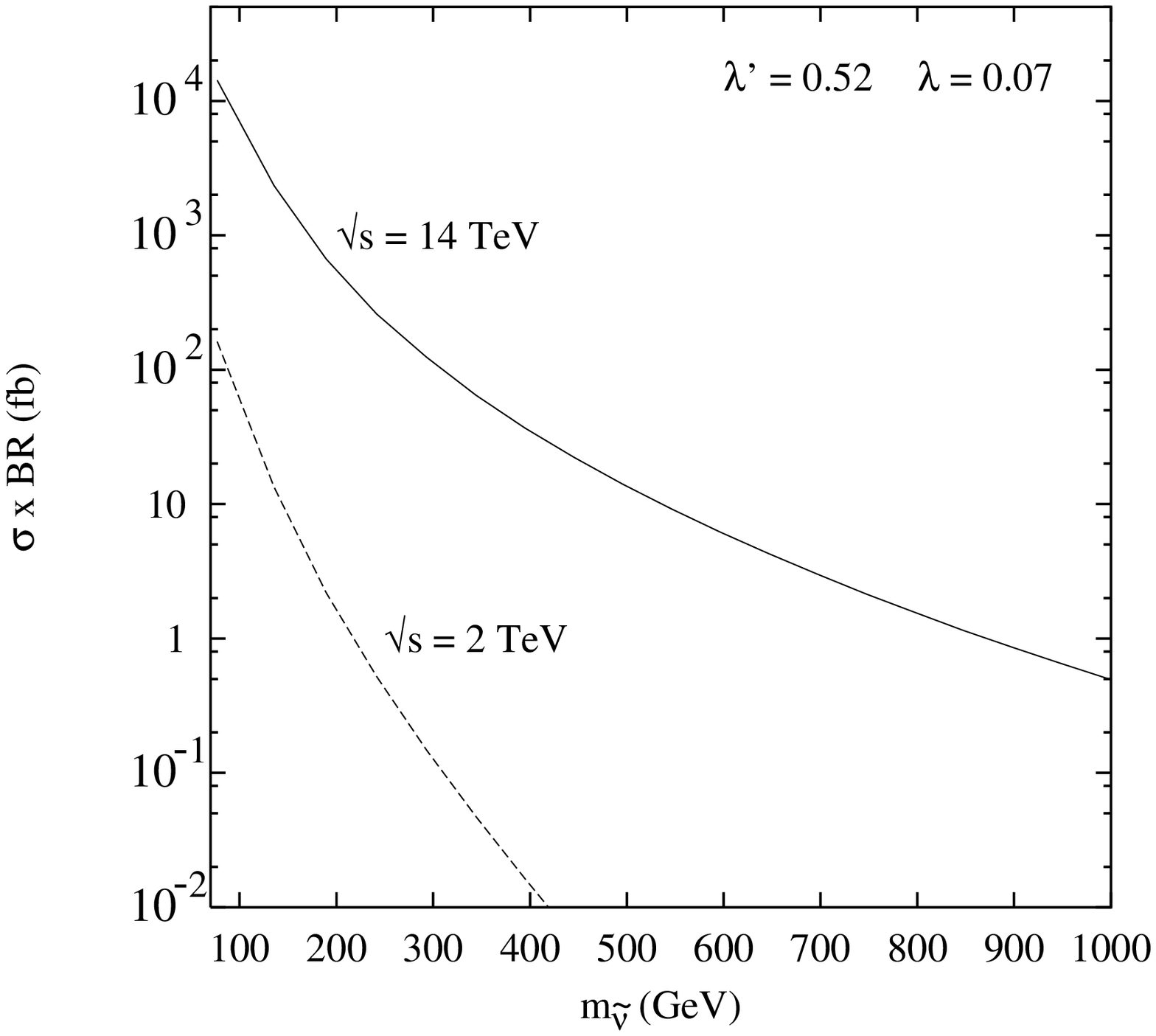}}
\mbox{\epsfxsize=7.truecm\epsfysize=9.5truecm\epsffile{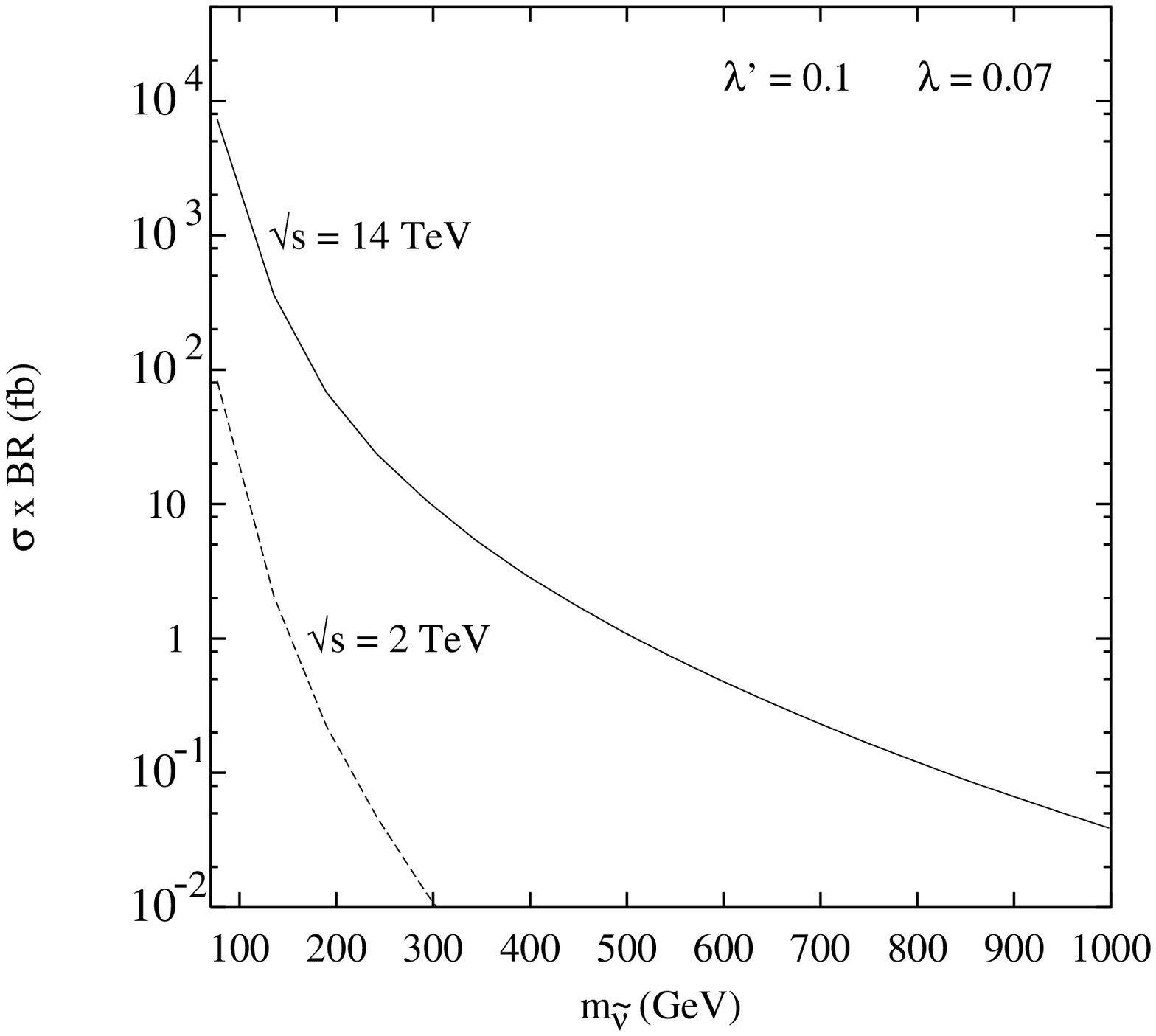}}
}
\caption{\label{mm}
The cross section for the process 
$pp \rightarrow bq\tilde\nu\rightarrow
bq \mu e$ as a 
function of the sneutrino mass with
$\lambda^{'}$
=0.52 (left figure) or 0.1 (right figure), and $\lambda$=0.07.
The solid line corresponds to LHC and the dashed one to Tevatron.}
\end{figure}

Another possibility for producing sneutrinos might be the diffractive
production \cite{KMR} which seems to offer an interesting
complementary way for Higgs production at the LHC.  We tried to make a
naive estimate of this possibility in the case of sneutrino.
Unfortunately a chirality flip in the quark loop would essentially
suppress the cross-section by the fourth power of the ratio of $b$ and
top quark masses, compared to the Higgs production cross section.
Even if the $R$-parity violating coupling were $\lambda'\sim 0.5$,
the cross section would be too small for the sneutrinos to be
detected.  Moreover, all the estimates of the Higgs production via
diffractive processes are plagued with huge uncertainties arising from
the non-perturbative aspects of QCD involved here. This huge systematic
error also affects the sneutrino production cross-section in a very
similar way. Considering the above two drawbacks, diffractive sneutrino
production at hadronic collision does not seem to be very promising.

Detection of muons at Tevatron and LHC experiments is straightforward.
Thus, even if a small $\lambda $ type coupling leading to a sneutrino
decaying to muons exists, the branching ratio may be big enough for
detecting a peak in the invariant mass.
In Fig. \ref{mm} we have considered a decay to a muon and an electron,
and plotted the $\sigma\times$BR for the values
of the couplings $\lambda^{'}$=0.52 and 0.1 and $\lambda_{321}$=0.07.
It is seen that the cross section at the LHC is above 1 fb up to
$m_{\tilde\nu}\sim 600$ GeV for $\lambda^{'}$=0.52 and
$m_{\tilde\nu}\sim 400$ GeV for $\lambda^{'}$=0.1.
Although the number of events at high masses is not large, the
narrow sneutrino peak should be easily detected.

\begin{figure}[t]
%\vspace*{-3truecm}
\centerline{
\mbox{\epsfxsize=7.truecm\epsfysize=7truecm\epsffile{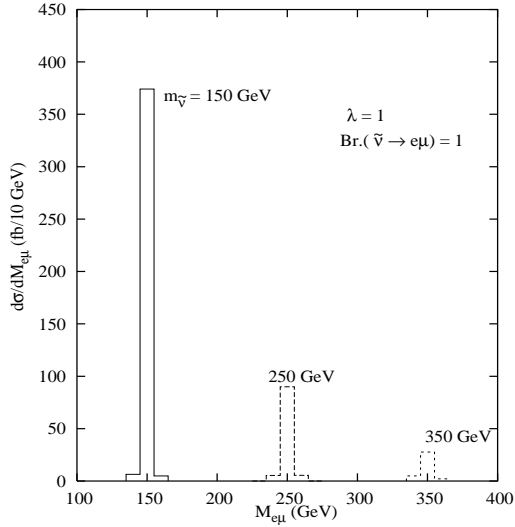}}
}
\caption{\label{mu_mas}
The $e \mu$-mass distribution in bins of 10 GeV.  
Here $\lambda_{321}$-coupling is assumed nonvanishing.}
\end{figure}

In Fig. \ref{mu_mas}, we present the $e \mu$-mass distribution in bins
of 10 GeV for three representative sneutrino masses. As we are
limited by the parton level analysis, we parametrise the
finite detector resolution and other effects due to e.m. radiation
by gaussian smearing the $p_T$of the $\mu$ and $e$, according to 
$\Delta p_T^l / p_T^l = 0.15/\sqrt{p_T^l} + 0.01$. We assume 100 \%
$\mu$- and $e$- detection efficiency.  Thus, there is essentially no physics
background to this flavour violating decay signal. For purpose of
illustration in Fig. \ref{mu_mas}, we have chosen the
$\lambda_{321}$-coupling to be non-zero as well as the sneutrino
decay branching ratio to $e \mu$ to be equal to 1.

As discussed earlier, in \cite{bews} it is proposed that the sneutrino
decay to photons could be used for detection.
Also, the rare scalar decay mode to two photons is
considered to offer the best possibility for detecting
the low-mass Higgs at the LHC.
In the case of the Higgs,
the relatively large branching ratio is due to the large top Yukawa
coupling.
In our case one might expect large branching
ratio, if the third generation $\lambda' $coupling is large.
However, the branching ratio is not large because of the 
chirality structure in the vertex.
In Ref. \cite{bews}, the
large production cross section compensates the small branching ratio,
and thus the $\gamma\gamma$ decay mode can be observable.
In our case, the cross section times branching ratio is less that
0.1 fb for sneutrinos heavier than 120 GeV, and thus we do not consider
this a viable search mode here.

As a summary, we conclude that the single production of sneutrino
in association with one or two $b$-jets may be observable at LHC
with sneutrino 
decaying to one or two $b$'s or to two leptons, if the $R_p$ violating 
couplings are not very small, but of the order of ${\cal{O}}(10^{-1})$.

\begin{flushleft} {\bf Acknowledgements} \end{flushleft}

\noindent
The authors thank the Academy of Finland
(project numbers 48787 and 54023) for financial support.
Z.-H. Yu thanks the World Laboratory, Lausanne, for the scholarship. 
The research of S.R. was partially supported by a Lady Davis postdoctoral
fellowship. A.D. acknowledges financial support from I.N.F.N., Sezione di
Roma.

\end{document}